\documentclass[a4paper]{jpconf}

\usepackage{iopams}
\usepackage{graphicx}
\usepackage{url}
\usepackage{hyperref}
\usepackage{physics}
\usepackage{xcolor}
\usepackage[normalem]{ulem}

\begin{document}

\hspace*{\fill} TIF-UNIMI-2023-7

\title{Towards an open-source framework to perform quantum calibration and characterization}

\author{A Pasquale$^{1,2}$, S Efthymiou$^1$, S Ramos-Calderer$^{1, 3}$, J Wilkens$^1$, I Roth$^1$ and S Carrazza$^{2,1,4}$}

\address{$^1$ Quantum Research Center, Technology Innovation Institute, Abu Dhabi, UAE.}
\address{$^2$ TIF Lab, Dipartimento di Fisica, Universit\`a degli Studi di Milano and INFN Sezione di Milano
Via Celoria 16, 20133, Milan, Italy.}
\address{$^3$ Departament de F\'isica Qu\`antica i Astrof\'isica and Institut de Ci\`encies del Cosmos (ICCUB), Universitat de Barcelona, Mart\'i i Franqu\`es 1, 08028 Barcelona, Spain.}
\address{$^4$ CERN, Theoretical Physics Department, CH-1211, Geneva 23, Switzerland.}

\begin{abstract}
    In this proceedings we present \texttt{Qibocal}, an open-source software package for calibration and characterization of quantum
    processing units (QPUs) based on the \texttt{Qibo} framework. \texttt{Qibocal} is specifically designed for self-hosted QPUs and
    provides the groundwork to easily develop, deploy and distribute characterization and calibration routines for all levels
    of hardware abstraction.
    \texttt{Qibocal} is based on a modular QPU platform agnostic approach and it provides a general purpose toolkit for superconducting qubits
    with the possibility of extensions to other quantum technologies.
    After motivating the need for such a module, we explain the program's flow and show examples of actual use
    for QPU calibration.
    We also showcase additional features provided by the library including automatic report generation
    and live plotting.

\end{abstract}

	\section{Introduction}
	The quantum computing landscape has rapidly evolved in recent years.
    Small noisy intermediate-scale quantum (NISQ)
    devices are starting to be accessible for in-house use in research groups and
	institutions.
    With that, the need for robust and general tools for qubit control has risen. Moreover,
	the characterization and calibration of the chip, invisible when accessing through cloud quantum computing
	providers, is an essential step to properly operate emerging quantum devices.

	We present \texttt{Qibocal}~\cite{qibocal_zenodo}, a Python module focused on the calibration and characterization of quantum devices.
	\texttt{Qibocal} is backed by both \texttt{Qibo}~\cite{Efthymiou_2021}, a library for the simulation of quantum circuits and
	systems, as well as \texttt{Qibolab}~\cite{stavros_efthymiou_2023_7748527}, a library built for quantum device control, which is able to transpile
	the logical operations of \texttt{Qibo} into instructions for the QPU.

	\texttt{Qibocal} introduces tools that support the calibration and characterization of QPUs on three different levels: \emph{development}, \emph{deployment} and \emph{distribution}.
    \texttt{Qibocal} provides
    a code library to rapidly \emph{develop} protocols for different hardware
    abstraction layers.
    The integration with \texttt{Qibo} allows one to easily switch
    between hardware execution and high-perfomance simulation \cite{qibojit}
    facilitating the theoretical study, numerical evaluation and hardware-optimization of protocols.
    Building on a library of characterization protocols, that is envisioned to rapidly grow, novel calibration and transpilation
    routines can take advantage of precise and detailed error and noise characterization.
    The protocols are \emph{deployed} using \emph{runcards}, which contain only the relevant information necessary to start calibration procedure.
    \texttt{Qibocal} is able to execute runcards containing protocols based on different abstraction layers making it
    possible for the user to execute full calibration pipelines, e.g.\ combining low-level characterization routines with gate-level performance
    benchmarks.
    The results of a runcard's protocols are collected
    in automatically generated human- and machine-readable reports.

    Finally, \texttt{Qibocal} is designed to encourage the \emph{distribution} of state-of-the-art calibration and characterization routines
    and their results. Reports can be stored in a self-hosted server archive, shared and further processed.
    Combined with the abstract, device-agnostic application programming interface (API) supporting different hardware setups, this architecture enables comprehensive comparative
    studies of hardware performance,  calibration and characterization protocols.

    In the following, we  briefly detail the role of \texttt{Qibocal} within the \texttt{Qibo} ecosystem, explain the workflow of a calibration
    and tools for retrieving results and sketch the capabilities in characterization of gate-sets before concluding.
    For further details and a technical documentation please refer to the official documentation available at:
    \begin{center}
        \url{https://qibo.science/qibocal/stable/}
    \end{center}

    \section{Qibocal within the Qibo framework}
    \begin{figure}
        \centering
        \includegraphics[width=0.7\textwidth]{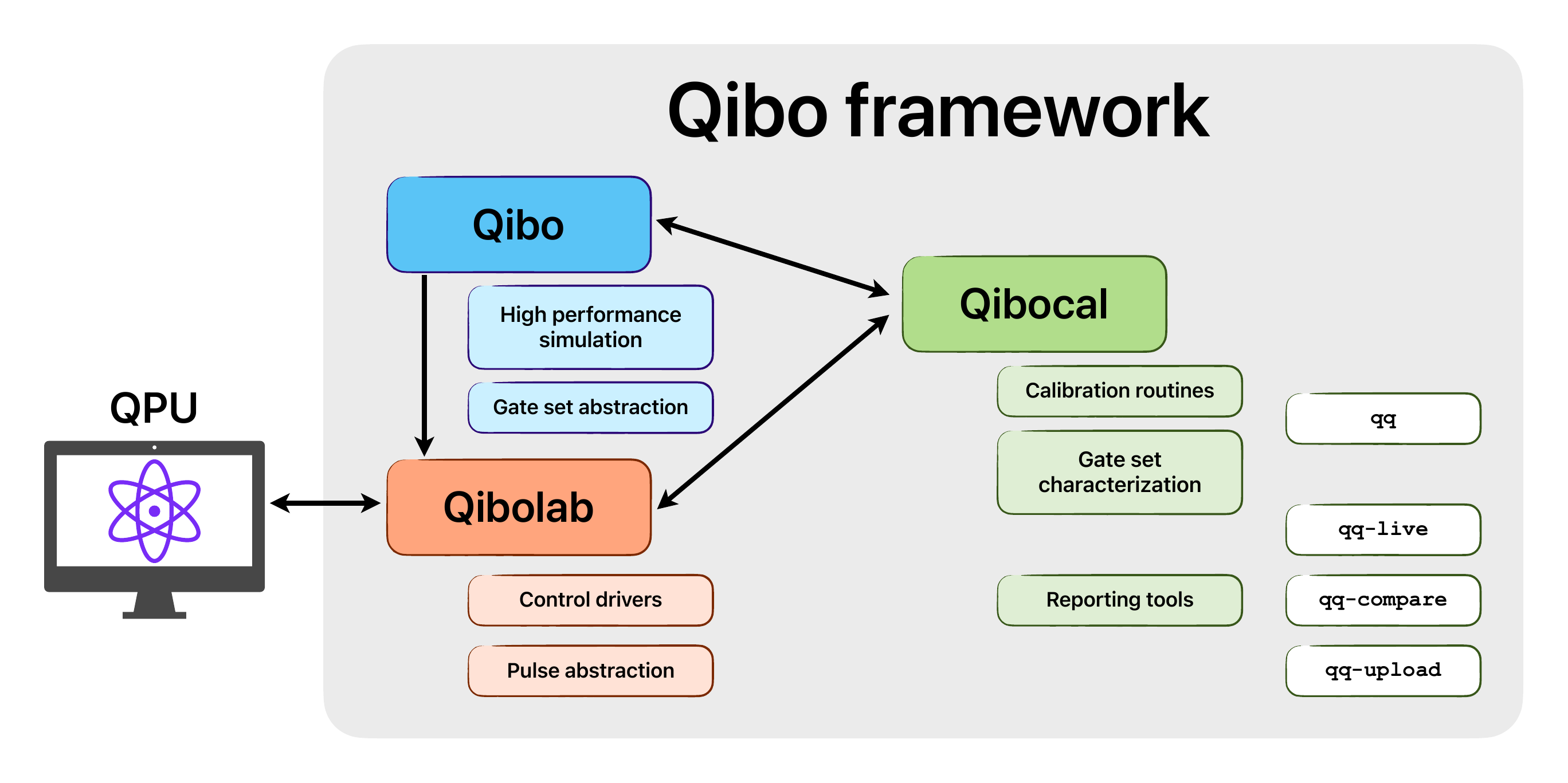}
        \caption{ \label{fig:flow} \texttt{Qibo} framework including \texttt{Qibocal} and \texttt{Qibolab}.}
    \end{figure}

    \texttt{Qibo} is evolving from a quantum circuit simulator  \cite{Efthymiou_2021, qibojit}
    to a full-stack software package to control quantum devices and calibrate QPUs.

    The transition between the simulation of a quantum circuit and its
    actual execution on QPUs is non-trivial,  given that quantum gates need to be promoted from
    linear algebra operations into controlled physical actions on a quantum system.
    In particular, the implementation of a gate on superconducting qubits is achieved by sending one or more microwave pulses
    that modify the state of the system. These pulses are generated through specific instruments that rely
    on different electronics.
    Within the \texttt{Qibo} framework, these low-level manipulations are achieved through \texttt{Qibolab}, a dedicated
    backend for the execution of quantum operations on hardware. The backend preserves the existing
    modularity of \texttt{Qibo} \cite{Qibo_proceeding}, and enables the user to switch from simulation to hardware
    execution with minimal effort.
    \texttt{Qibolab} provides an abstraction layer that is able to operate platforms
    with different control electronics using the same API. This \emph{platform agnostic} approach
    enables one to accommodate different hardware setups while reducing the effort needed to maintain the library.

    Full control on the electronics in itself is of course not enough to run a quantum computer.
    The reason being that quantum hardware, particularly those based on superconducting technologies, require precise fine-tuning and \emph{calibration} of the parameters of the control waveforms.
    In particular, mis-calibration of one of these parameters leads to noisy and incorrect results.
    The calibration parameters comprise physical quantities, such as the qubit or resonator frequencies, as well as control
    variables, such as the $\pi$-pulse amplitude that induces a transition between the ground 
    and excited 
    state of the qubit.

    Suitable parameter values are determined through a set of experiments which we call \emph{calibration routines}.
    On the level of pulse control, each calibration routine aims at fine-tuning one or more parameters of the waveforms.
    For example, in the case of a single qubit coupled to a resonator cavity the first calibration routine involves a resonator spectroscopy, that aims at finding the frequency of the resonator by sweeping a frequency range.

    To unlock the full potential of NISQ devices, more complex calibration routines can further infer a precise characterization of coherent errors and noise sources or evaluate the devices performance in high-level tasks.
    Such information can be used, e.g.\ for further improving the pulse control, adapt the transpilation, and inform the interpretation of read-out.

    From a software perspective,
    this motivates the development of a general purpose toolkit that is able to collect and develop
    all routines to calibrate QPUs and provides a high-degree of automation for scheduling and executing pipelines of multiple calibration routines.
    This is precisely why we introduce \texttt{Qibocal}.
    In particular, by being a separate module within the \texttt{Qibo} framework, it can define calibration pipelines that combine protocols using the different abstraction layers provide by \texttt{Qibo} and \texttt{Qibolab}, as illustrated in Fig.~\ref{fig:flow}.

    By the \emph{platform agnostic} layout of \texttt{Qibolab}, the calibration routines available in
    \texttt{Qibocal} are independent on the specifics of the platform and the related instruments.
    This enables developers and experimentalists to focus specifically on the development of
    calibration routines, delegating the hardware-level control to \texttt{Qibolab},
    without introducing code duplication.
    Using the gate-layer abstraction of \texttt{Qibo}, calibration protocols can define experiments in terms of gate circuits and compare to classical simulations.

    In the next section we explain how calibration routines are deployed to calibrate a quantum processor with \texttt{Qibocal}.

    \begin{figure}
        \centering
        \includegraphics[width= 0.8 \textwidth]{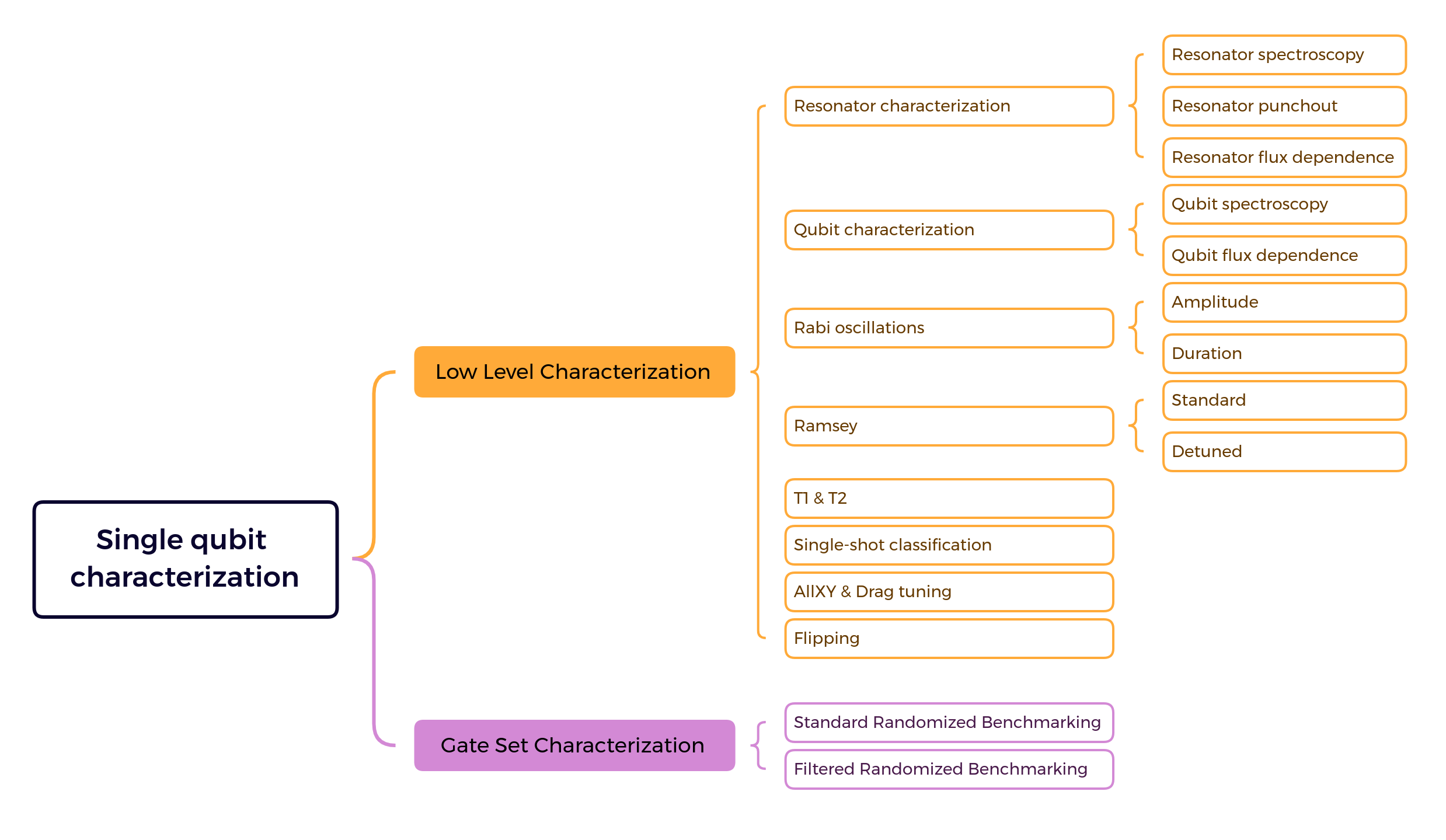}
            \caption{\label{fig:calibration}Current single-qubit routines available in \texttt{Qibocal}.
            }
    \end{figure}

	\section{How to calibrate a quantum device}

    In \texttt{Qibocal} we adopt a declarative programming paradigm \cite{abe0560f66034ae3893128ebc0e8197e}
    to quantum calibration. This approach focuses on what the user would like to do without requiring
    any a priori knowledge on how the code is executed. Our aim is to streamline the
    characterization of quantum processors that often involves complicated measurements
    \cite{automatic} that have a non-trivial scheme of mutual dependencies \cite{optimus}.

    All the information necessary to start calibrating a quantum device is encoded in a single file,
    which we call \emph{runcard}\footnote{For more information on how to write a runcard we invite the interested
    reader to consult the \texttt{Qibocal} documentation \url{https://qibo.science/qibocal/stable/} .}.
    This runcard generally contains a sequence of routines or protocols to be executed and information regarding which
    platform, and which qubits within that platform, will be calibrated as well as the format in which the data acquired will be stored.

    In Fig.~\ref{fig:calibration} we give an overview of the most important calibration routines that are currently available in version
    \texttt{0.0.1} of \texttt{Qibocal} and can be invoked by a runcard. In this first release, we provide all the experiments necessary to characterize superconducting {single-}qubit devices
    starting from the characterization of the resonator, up to single shot classification.
    Moreover, we provide routines to compute the decoherence times, $T_1$ and $T_2$, of the qubit as well as the
    abstractions necessary for Randomized Benchmarking protocols to extract fidelities.

    To launch a calibration script we provide the command \texttt{qq}, that followed by the path to the runcard,

    \begin{center}
        \texttt{qq} $<$\texttt{path\_to\_runcard}$>$,
    \end{center}
    executes the queued calibration routines and generate a report of the results.
    At the end of the program the user is able to collect the following: data from the executed
    routines in the format specified in the runcard, the determined calibration parameters
    and an \texttt{html} report containing a graphical summary of the results. 
    When multiple actions are specified in the runcard they are executed sequentially and all the
    parameters tuned by one calibration routine are fed to the next one. In this way, we are
    able to calibrate different parameters by passing to the platform the parameters tuned on the fly.

    Moreover, all the coded routines are compatible with multiplexed readout \cite{multiplex}, if available,
     which allows for the measurement of multiple qubits at the same time with high fidelity. In this way we are able to calibrate more than one qubit
    at a time by executing the same characterization routine on multiple qubits simultaneously.
    \begin{figure}
        \centering
        \includegraphics[width=0.8 \textwidth]{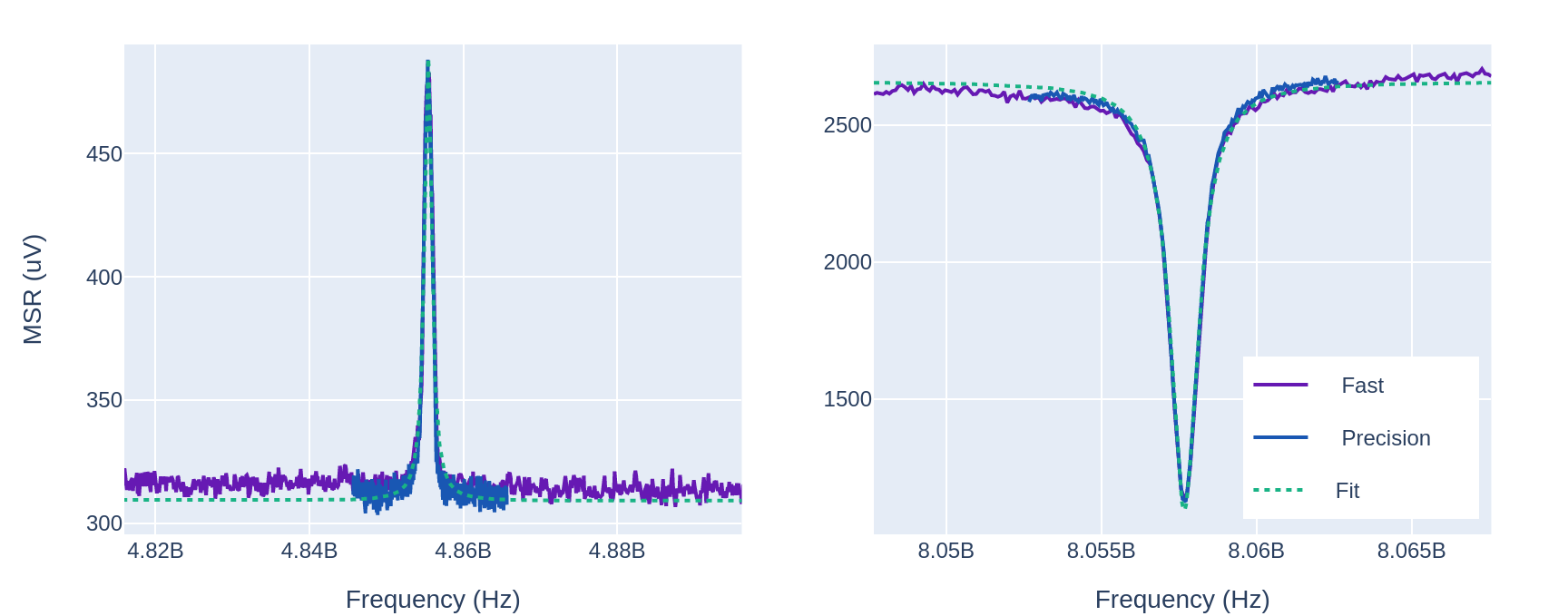}
        \caption{ \label{figure:plots} Example of two calibration experiments executed using \texttt{Qibocal} on quantum processors.
        On the left a qubit spectroscopy and on the right a resonator spectroscopy for a qubit hosted at Quantum Reasearch Center (QRC)
        in Technology Innovation Institute (TII).}
    \end{figure}

    \section{Non-interactive gate-set characterization}
    The calibration and operation of NISQ devices requires
    characterization protocols assessing the device's performance on
    different layers of abstraction, for example, using the average performance in implementing
    quantum gates from a group \cite{EggerWilhelm:2014,KellyEtAl:2014}.

    As a first step towards providing a large collection of characterization protocols for further
    layers of abstraction, \texttt{Qibocal} implements an API to quickly define \emph{non-interactive
    gate-set characterization} protocols.
    This class of protocols comprises inter alia, randomized benchmarking schemes \cite{HelsenEtAl:2022}, quantum state, process
    and detector certification \cite{KlieschRoth:2021} and tomography \cite{ChuangNieslen:1997,
    ParisRehacek:2004} protocols as well as robust \cite{PhysRevX.4.011050,FlammiaWallman:2020,HelsenEtAl:2021} and self-consistent variants \cite{Nielsen:2021}.
    The experiment of a (non-interactive) gate-set characterization protocol is defined in terms of a
    (potentially random) ensemble of circuits from which output statistics are collected.
    The collected data is post-processed by first performing a sequence of calculations on the measured data of each circuit and, then, calculating estimates based on the aggregated statistics over different circuits.

    To showcase the API, the current version provides an implementation of standard randomized benchmarking for single qubit Clifford gates \cite{Magesan:2010} as well as filtered randomized benchmarking for different smaller gate sets \cite{HelsenEtAl:2022}.

    \section{Reporting tools}
    \subsection{Report generation and comparison}
\begin{figure}
        \includegraphics[width=\textwidth]{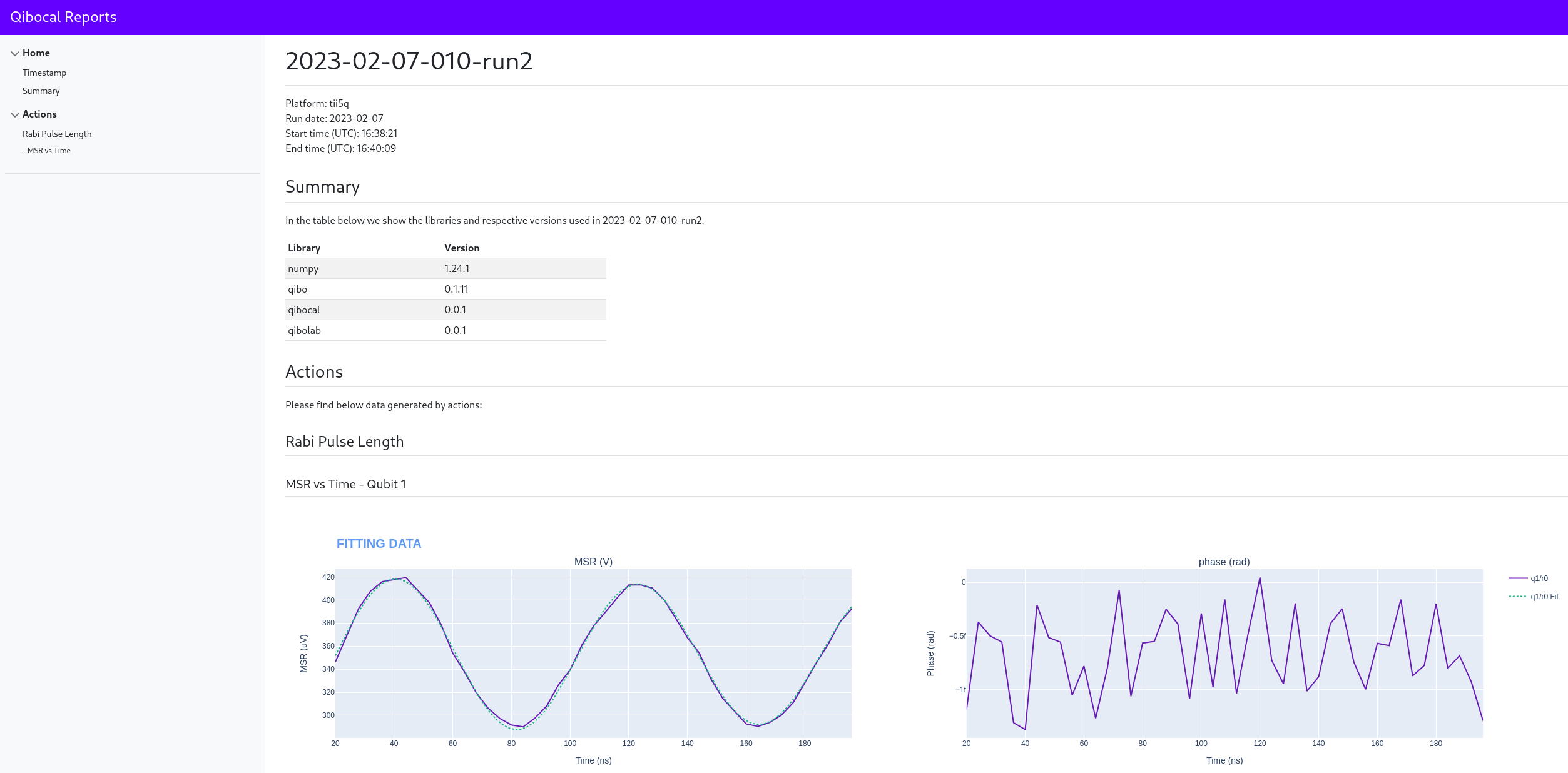}
        \caption{\label{fig:report} Example of a report generated using \texttt{Qibocal} for a quibt hosted at QRC in TII.}
    \end{figure}

    An example of a report generated by \texttt{qq} is shown in Fig.~\ref{fig:report}.
    The report includes general information about the experiment, versions of dependencies used and
    plots for all the routines executed.

    The reports obtained after executing a calibration can be uploaded to a web page that serves as an
    archive for collecting results for calibration parameters. We provide the dedicated command, \texttt{qq-upload}
    for this purpose. This feature is particularly useful as it enables experimentalists to share
    the latest parameters found with ease.

    We also offer the possibility to compare two or more reports together through the \texttt{qq-compare}
    command. During the comparison, a new report is generated that contains data from all the combined reports.
    In the context of quantum calibration this is, for example, useful to monitor the drift of optimal parameters with time \cite{drifts,drifts2}.

    \subsection{Live-plotting}
     From an experimental point of view, it is valuable to visualize the data as it is acquired
     in real time. It enables experimentalists to detect and fix possible issues in real time, without
     the need to wait until the end of the execution of the program, which may require a long time or manually interventions.
     This is for example often relevant for the first stages in the calibration of experimental superconducting qubits where
     scans over wide ranges of parameters are required.
     \texttt{Qibocal} supports \textit{live-plotting} through the command \texttt{qq-live},
     which enables the user to see live updates directly from the QPU which is currently under calibration.

    On top of this, we also provide \textit{live-fitting} which consists in analyzing the data
    as it as acquired. In particular, alongside the raw data the user will be able to see an estimate
    of the fitted parameters and the relative curve as shown in Fig.~\ref{figure:plots}.

    \section{Conclusion and outlook}
    In this proceedings we have presented \texttt{Qibocal}, a package within the \texttt{Qibo}
    framework to calibrate and characterize quantum devices.
    \texttt{Qibocal} is one of the first fully open-source
    Python libraries for quantum calibration.
    It has already been used to
    calibrate superconducting chips with up to 5 flux-tunable transmon qubits.
    The code is available at:
    \begin{center}
        \url{https://github.com/qiboteam/qibocal}
    \end{center}

    In the future we are looking to extend the capabilities of \texttt{Qibocal} in multiple directions:
    First, we aim at providing the means for calibrating multi-qubit gates and develop calibration
    routines to be able to control entangled qubits consistent with the state-of-the-art \cite{Rol_2020, Sung}.

    Second, we plan to provide more flexible automatization of calibration pipelines going beyond the current sequential execution of multiple routines.
    One focus here is on reducing the time required for re-calibration of a QPU, crucial for systems with a large number of qubits.

    Third, together with collaborators, we pursue the implementation of a comprehensive set of characterization protocol for gate implementations, cross-talk and read-out.

    Finally, using the growing capabilities of \texttt{Qibocal}, we hope to in the future be able to perform comparative studies of different approaches to quantum hardware implementations and calibration.

\section*{References}
\bibliographystyle{iopart-num} 
\bibliography{ref} 

\providecommand{\newblock}{}
\begin{thebibliography}{10}
\expandafter\ifx\csname url\endcsname\relax
  \def\url#1{{\tt #1}}\fi
\expandafter\ifx\csname urlprefix\endcsname\relax\def\urlprefix{URL }\fi
\providecommand{\eprint}[2][]{\url{#2}}

\bibitem{qibocal_zenodo}
Pasquale A, Sarlle D, Efthymiou S, Carrazza S, Pedicillo E, Orgaz A, Sopena A,
  Robbiati M, Hantute M, Jadwiga W and \emph{et al} 2023 qiboteam/qibocal:
  Qibocal 0.0.1 \urlprefix\url{https://doi.org/10.5281/zenodo.7662185}

\bibitem{Efthymiou_2021}
Efthymiou S, Ramos-Calderer S, Bravo-Prieto C, P{\'{e}}rez-Salinas A,
  Garc{\'{\i}}a-Mart{\'{\i}}n D, Garcia-Saez A, Latorre J~I and Carrazza S 2021
  {\em Quantum Science and Technology\/} {\bf 7} 015018

\bibitem{stavros_efthymiou_2023_7748527}
Efthymiou S, Orgaz A, Carrazza S, Pasquale A, Lazzarin M, Sarlle D, Bordoni S,
  Pedicillo E, Paul, Hantute M, Candido A, Pérez J, atomicprinter, Robbiati M,
  arckpx, qrcworkshop and Kolesnyk D 2023 qiboteam/qibolab: Qibolab 0.0.2
  \urlprefix\url{https://doi.org/10.5281/zenodo.7748527}

\bibitem{qibojit}
Efthymiou S, Lazzarin M, Pasquale A and Carrazza S 2022 {\em Quantum\/} {\bf 6}
  814

\bibitem{Qibo_proceeding}
Carrazza S, Efthymiou S, Lazzarin M and Pasquale A 2023 {\em Journal of
  Physics: Conference Series\/} {\bf 2438} 012148

\bibitem{abe0560f66034ae3893128ebc0e8197e}
Lloyd J 1994 Practical advantages of declarative programming {\em Unknown\/} pp
  3 -- 17 conference Proceedings/Title of Journal: Joint Conference on
  Declarative Programming

\bibitem{automatic}
Xu Y, Huang G, Balewski J, Morvan A, Nowrouzi K, Santiago D, Naik R, Mitchell B
  and Siddiqi I 2022 {\em ACM Transactions on Quantum Computing\/} {\bf 4}

\bibitem{optimus}
Kelly J, O'Malley P, Neeley M, Neven H and Martinis J~M 2018 Physical qubit
  calibration on a directed acyclic graph

\bibitem{multiplex}
Heinsoo J, Andersen C~K, Remm A, Krinner S, Walter T, Salath{\'{e} } Y,
  Gasparinetti S, Besse J~C, Poto{\v{c}}nik A, Wallraff A and Eichler C 2018
  {\em Physical Review Applied\/} {\bf 10}

\bibitem{EggerWilhelm:2014}
Egger D~J and Wilhelm F~K 2014 {\em Phys. Rev. Lett.\/} {\bf 112}(24) 240503

\bibitem{KellyEtAl:2014}
Kelly J, Barends R, Campbell B, Chen Y, Chen Z, Chiaro B, Dunsworth A, Fowler
  A~G, Hoi I~C, Jeffrey E, Megrant A, Mutus J, Neill C, O'Malley P~J~J,
  Quintana C, Roushan P, Sank D, Vainsencher A, Wenner J, White T~C, Cleland
  A~N and Martinis J~M 2014 {\em Phys. Rev. Lett.\/} {\bf 112}(24) 240504

\bibitem{HelsenEtAl:2022}
Helsen J, Roth I, Onorati E, Werner A and Eisert J 2022 {\em PRX Quantum\/}
  {\bf 3}(2) 020357

\bibitem{KlieschRoth:2021}
Kliesch M and Roth I 2021 {\em PRX Quantum\/} {\bf 2}(1) 010201

\bibitem{ChuangNieslen:1997}
Chuang I~L and Nielsen M~A 1997 {\em Journal of Modern Optics\/} {\bf 44}
  2455--2467

\bibitem{ParisRehacek:2004}
Paris M and Rehacek J 2004 {\em Quantum state estimation\/} vol 649 (Springer
  Science \& Business Media)

\bibitem{PhysRevX.4.011050}
Kimmel S, da~Silva M~P, Ryan C~A, Johnson B~R and Ohki T 2014 {\em Phys. Rev.
  X\/} {\bf 4}(1) 011050

\bibitem{FlammiaWallman:2020}
Flammia S~T and Wallman J~J 2020 {\em ACM Transactions on Quantum Computing\/}
  {\bf 1} 1--32

\bibitem{HelsenEtAl:2021}
Helsen J, Ioannou M, Kitzinger J, Onorati E, Werner A, Eisert J and Roth I 2021
  {\em arXiv preprint arXiv:2110.13178\/}

\bibitem{Nielsen:2021}
Nielsen E, Gamble J~K, Rudinger K, Scholten T, Young K and Blume-Kohout R 2021
  {\em {Quantum}\/} {\bf 5} 557 ISSN 2521-327X

\bibitem{Magesan:2010}
Magesan E, Gambetta J and Emerson J {\em arXiv preprint arXiv:1009.3639\/}

\bibitem{drifts}
Koch C~P, Palao J~P, Kosloff R and Masnou-Seeuws F 2004 {\em Physical Review
  A\/} {\bf 70}

\bibitem{drifts2}
Kelly J, Barends R, Fowler A~G, Megrant A, Jeffrey E, White T~C, Sank D, Mutus
  J~Y, Campbell B, Chen Y, Chen Z, Chiaro B, Dunsworth A, Hoi I~C, Neill C,
  O'Malley P~J~J, Quintana C, Roushan P, Vainsencher A, Wenner J, Cleland A~N
  and Martinis J~M 2015 {\em Nature\/} {\bf 519} 66--69

\bibitem{Rol_2020}
Rol M, Ciorciaro L, Malinowski F, Tarasinski B, Sagastizabal R, Bultink C,
  Salathé Y, Haandbaek N, Sedivy J and DiCarlo L 2020 {\em Applied Physics
  Letters\/} {\bf 116} 054001

\bibitem{Sung}
Sung Y, Ding L, Braumüller J, Vepsäläinen A, Kannan B, Kjaergaard M, Greene
  A, Samach G, McNally C, Kim D, Melville A, Niedzielski B, Schwartz M, Yoder
  J, Orlando T, Gustavsson S and Oliver W 2021 {\em Physical Review X\/} {\bf
  11}

\end{thebibliography}
\end{document}